\documentclass[prl,aps,floats,amsmath,amssymb,twocolumn,superscriptaddress,preprintnumbers,nofootinbib]{revtex4}
\usepackage{graphicx}
\usepackage{dcolumn}
\usepackage{bm}
\usepackage{hyperref}
\usepackage{epstopdf}
\usepackage{color}
\usepackage{hyperref}
\usepackage{epsfig,latexsym,cancel,amssymb,amsmath}
\graphicspath{{./Figures/}}

\def\lsim{\raisebox{-4pt}{$\,\stackrel{\textstyle{<}}{\sim}\,$}}
\def\gsim{\raisebox{-4pt}{$\,\stackrel{\textstyle{>}}{\sim}\,$}}

\def\beq{\begin{equation}} 
\def\eeq{\end{equation}} 
\def\bea{\begin{eqnarray}} 
\def\eea{\end{eqnarray}} 
\def\ben{\begin{enumerate}} 
\def\een{\end{enumerate}}

\def\lsim{\mathrel{\raise.3ex\hbox{$<$\kern-.75em\lower1ex\hbox{$\sim$}}}} 
\def\gsim{\mathrel{\raise.3ex\hbox{$>$\kern-.75em\lower1ex\hbox{$\sim$}}}} 
\def\ifmath#1{\relax\ifmmode #1\else $#1$\fi}

\newcommand{\met}{\mbox{${\not\! E}_{\rm T}$}}

\begin{document}
\DeclareGraphicsExtensions{.jpg,.pdf,.mps,.png,}

\preprint{FERMILAB-PUB-13-186-T}
\preprint{LA-UR-13-24173}
\title{Supersymmetric Exotic Decays of the 125 GeV Higgs Boson}

\author{Jinrui Huang}
\affiliation{Theoretical Division, T-2, MS B285,
Los Alamos National Laboratory, Los Alamos, NM 87545, USA}

\author{Tao Liu}
\affiliation{Department of Physics, The Hong Kong University of
  Science and Technology, Clear Water Bay, Kowloon, Hong Kong}

\author{Lian-Tao Wang}
\affiliation{Enrico Fermi Institute, 
University of Chicago, Chicago, IL 60637, USA}
\affiliation{KICP and Dept. of Physics, University of Chicago, 5640
  S. Ellis Ave., Chicago, IL 60637, USA}

\author{Felix Yu}
\affiliation{Theoretical Physics Department, Fermi National
  Accelerator Laboratory, P.~O.~Box 500, Batavia, IL 60510, USA}


\begin{abstract}

We reveal a set of novel decay topologies for the 125 GeV Higgs boson
in supersymmetry which are initiated by its decay into a pair of
neutralinos, and discuss their collider search strategies.  This
category of exotic Higgs decays are characterized by the collider
signature: visible objects + $\met$, with $\met$ dominantly arising
from escaping dark matter particles.  Their benchmark arises naturally
in the Peccei-Quinn symmetry limit of the MSSM singlet-extensions,
which is typified by the co-existence of three light particles:
singlet-like scalar $h_1$ and pseudoscalar $a_1$, and singlino-like
neutralino $\chi_1$, all with masses of $\lesssim 10$ GeV, and the
generically suppression of the exotic decays of the 125 GeV Higgs
boson $h_2\to h_1 h_1$, $a_1a_1$ and $\chi_1\chi_1$, however.  As an
illustration, we study the decay topology: $h_2 \to \chi_1 \chi_2$,
where the bino-like $\chi_2$ decays to $h_1 \chi_1$ or $a_1 \chi_1$,
and $h_1/a_1 \to f\bar f$, with $f\bar f = \mu^+\mu^-$, $b\bar b$.  In
the di-muon case ($m_{h_1/a_1} \sim 1$ GeV), a statistical sensitivity
of $\frac{S}{\sqrt{B}} > 6 \sigma$ can be achieved easily at the 8 TeV
LHC, assuming $\frac{\sigma(pp \rightarrow W h_2)}{\sigma(pp
  \rightarrow W h_{\rm SM})} {\rm Br}(h_2 \to \mu^+\mu^- \chi_1
\chi_1)=0.1$. In the $b\bar b$ case ($m_{h_1/a_1} \sim 45$ GeV), 600
fb$^{-1}$ data at the 14 TeV LHC can lead to a statistical sensitivity
of $\frac{S}{\sqrt{B}} > 5 \sigma$, assuming $\frac{\sigma(pp
  \rightarrow Z h_2)}{\sigma(pp \rightarrow Z h_{\rm SM})} {\rm
  Br}(h_2 \to b\bar b \chi_1 \chi_1)=0.5$.  These exotic decays open a
new avenue for exploring new physics couplings with the 125 GeV Higgs
boson at colliders.
\end{abstract}

\maketitle

{\bf [Introduction]} The discovery of a 125 GeV Higgs resonance at the
CMS~\cite{:2012gu} and ATLAS~\cite{:2012gk} experiments has launched
an era of precision Higgs phenomenology, emphasizing $CP$ and spin
discrimination and exotic Higgs decays.  Of particular interest are
exotic Higgs decays that arise in well-motivated new physics (NP)
scenarios aimed at solving the gauge hierarchy problem by stabilizing
the Higgs mass against divergent quantum corrections, such as
supersymmetry (SUSY). The Higgs mass stabilization mechanism generally
manifests itself through Higgs couplings absent in the Standard Model
(SM). The 125 GeV Higgs therefore is expected to be a leading window
into NP.

Because the 125 GeV SM Higgs decay width is small ($\Gamma \sim 4$
MeV), a new coupling between the Higgs boson and some light particles
may lead to a large exotic Higgs decay branching fraction.  The
current bounds on such channels are still weak: a branching fraction
as large as $\sim 60\%$ is allowed at the $2\sigma$
C.L.~\cite{Belanger:2013kya, Giardino:2013bma, Djouadi:2013qya,
  ATLAS:2013sla, CMS:yva}, in a general context, e.g., if new physics
is allowed to enter the Higgs-glue-glue coupling.  If SM couplings are
assumed, theorist-performed fits constrain the invisible branching
fraction to be $\lesssim 25\%$ at 95\% C.L.~\cite{Belanger:2013kya,
  Giardino:2013bma}.  Even with the full 300 fb$^{-1}$ of the 14 TeV
LHC, the projected upper bound is $\sim 10\%$ at the $2\sigma$ C.L. on
such channels~\cite{Peskin:2012we} (mainly driven by estimates of
systematic errors), which still leaves appreciable room for an exotic
decay mode.  Searches for exotic decays are therefore very natural and
effective tools to explore possible and exciting new couplings to the
125 GeV Higgs boson.

Exotic Higgs decays are often grouped into two categories according to
their collider signatures: (1) purely $\met$; (2) visible objects (no
$\met$, except for neutrinos from heavy quark or tau decays).  Case
(1) is mainly dark matter (DM) motivated and was originally studied in~\cite{Gunion:1993jf}.  A well-known
example for case (2) is the $R$-symmetry limit of the
Next-to-Minimal Supersymmetric Standard Model
(NMSSM)~\cite{Dobrescu:2000jt}, in which the SM-like Higgs can
significantly decay to a pair of light singlet-like $R$-axions
($a_1$).

Separately, there is the deep cosmic mystery of DM.  In
the past decade, a few DM direct detections have reported excesses,
which can controversially be interpreted as hints of a sub-electroweak
(EW) scale DM particle with a relatively large spin-independent direct
detection cross section~\cite{expts1, expts2, expts3, expts4}. One of
the most interesting possibilities arises from the approximate
Peccei-Quinn (PQ) symmetry limit in the NMSSM~\cite{Draper:2010ew,
  Kozaczuk:2013spa}.  In this scenario, the lightest neutralino is
singlino-like and has a mass of $\lesssim 10$ GeV, providing a
naturally light DM candidate.  Moreover, pair annihilation into the
light pseudoscalar as well as exchange of the light scalar with
nucleons allow the singlino to achieve simultaneously the correct
relic density and the large direct detection cross section indicated
by some experiments~\cite{Draper:2010ew}.

In this letter we will note that the PQ-symmetry limit not only
provides a supersymmetric benchmark for sub-EW scale DM, but we also
emphasize its very rich Higgs physics.  The phenomenology is
characterized by new SM-like Higgs ($h_2$) exotic decays: $h_2 \to
\chi_1 \chi_2, \chi_2\chi_2$, with the subsequent decays $\chi_2 \to
h_1 \chi_1$, $a_1 \chi_1$ and $h_1$, $a_1$ into SM particles (here
$\chi_1$ and $\chi_2$ are the lightest and the second lightest
neutralinos, respectively).  The PQ-symmetry limit therefore provides
benchmarks for a third category of exotic Higgs decays: (3) visible
objects and $\met$, where $\met$ dominantly arises from DM candidates.

{\bf [Theoretical Motivations]} To address the notorious $\mu$ problem
in the MSSM, various singlet-extensions of the MSSM have been
explored, such as the NMSSM~\cite{NMSSM} and the nearly-MSSM
(nMSSM)~\cite{nMSSM}, where different symmetries are introduced to
forbid the bare $\mu$ term in the MSSM. These models share a common
global PQ-symmetry limit~\cite{Barger:2006dh}, with the superpotential
and soft SUSY-breaking terms given by
\begin{align}
\mathbf{W} &= \lambda \mathbf{S} \mathbf{H_u} \mathbf{H_d} +{\mathcal O}(\kappa) \ , 
\nonumber \\
V_{\text{soft}} &= {m^2_{H_d}} |H_d|^2 + {m^2_{H_u}} |H_u|^2 + {m^2_S}|S|^2 
\nonumber \\
 &- (\lambda A_{\lambda} H_u H_d S + \text{ h.c.} ) +{\mathcal O}(\kappa) \ ,
\label{eqn:PQlimit}
\end{align}
where $H_d$, $H_u$ and $S$ are the neutral fields of the
${\bf H_d}$, ${\bf H_u}$ and ${\bf S}$ superfields, respectively.
Small explicit PQ-breaking terms (denoted by ${\mathcal O}(\kappa)$), such as $\kappa {\bf S}^3$ and its
softly SUSY breaking term $A_\kappa \kappa S^3$ in the NMSSM, are surely allowed in
realistic scenarios.  Once the singlet scalar $S$ obtains a vacuum
expectation value (VEV) $\langle S \rangle = v_S$, an effective $\mu$ 
parameter $\mu = \lambda v_S$ can be generated.  Since the current
LHC data constrains large mixing between the $CP$-even and
$CP$-odd Higgs sector~\cite{Djouadi:2013qya, Shu:2013uua}, we assume
no $CP$-violation in the Higgs sector.

One feature of this scenario is that in the decoupling limit ($\lambda
=\frac{\mu }{ v_S} \lesssim \mathcal{O}(0.1)$), the lightest $CP$-even
($h_1$), $CP$-odd ($a_1$) Higgs bosons and the lightest neutralino
($\chi_1$) form an approximate singlet-like PQ-axion supermultiplet.
These states, mainly the saxion, axion, and axino, respectively, are
much lighter than the PQ-symmetry breaking scale.  This is because the
PQ symmetry breaking is mainly controlled by the singlet superfield
${\bf S}$ in this scenario, while the mass splittings among the
saxion, axion and axino are induced by SUSY breaking and are
suppressed (recall, if SUSY is unbroken, their masses are degenerate).
Explicitly, $m_{h_1}^2$, $m_{\chi_1}$ are given by
\bea 
m_{h_1}^2 &=& - 4 v^2 \varepsilon^2 + 
\frac{4 \lambda^2 v^2 }{ \tan^2
  \beta} \left (1- \frac{ \varepsilon m_Z}{\lambda \mu} \right)
  \left (1+ \frac{ 2 \varepsilon \mu}{\lambda m_Z}
  \right)
\nonumber \\ 
&+& 
16 \frac{v^4}{m_Z^2} \varepsilon^4 + 
\sum\limits_{i=0}^{5} \mathcal{O} \left( \frac{
  \lambda^{5-i}}{\tan^i \beta}, \kappa \right) \ , \nonumber \\ 
m_{\chi_1}& =& -\frac{\lambda^2 v^2 \sin 2\beta}{\mu} +
\sum\limits_{i=0}^{5} \mathcal{O} \left(\frac{\lambda^{5-i}}{ \tan^i \beta}
, \kappa \right) , 
\eea
with $\varepsilon = \frac{\lambda \mu}{m_Z} \varepsilon'$,
$\varepsilon' = \left( \frac{A_\lambda}{\mu \tan \beta } - 1\right)$.
Avoiding a tachyonic $h_1$ mass immediately requires
\bea
\varepsilon^2 < \frac{\lambda^2}{\tan^2 \beta} + {\rm loop  \ corrections} \ . 
\eea
This constraint has important implications for the decay of the SM-like Higgs.
Note the contribution of $Z \to \chi_1 \chi_1$ to the $Z$ invisible
decay width is small, because the non-singlino content in $\chi_1$ is
of the order $\lambda v / \mu$, and the $Z \chi_1 \chi_1$ coupling is
suppressed by $(\lambda v / \mu)^2$, where $v = 174$ GeV.

\begin{figure*}[ht]
\begin{center}
\includegraphics[width=0.24\textwidth]{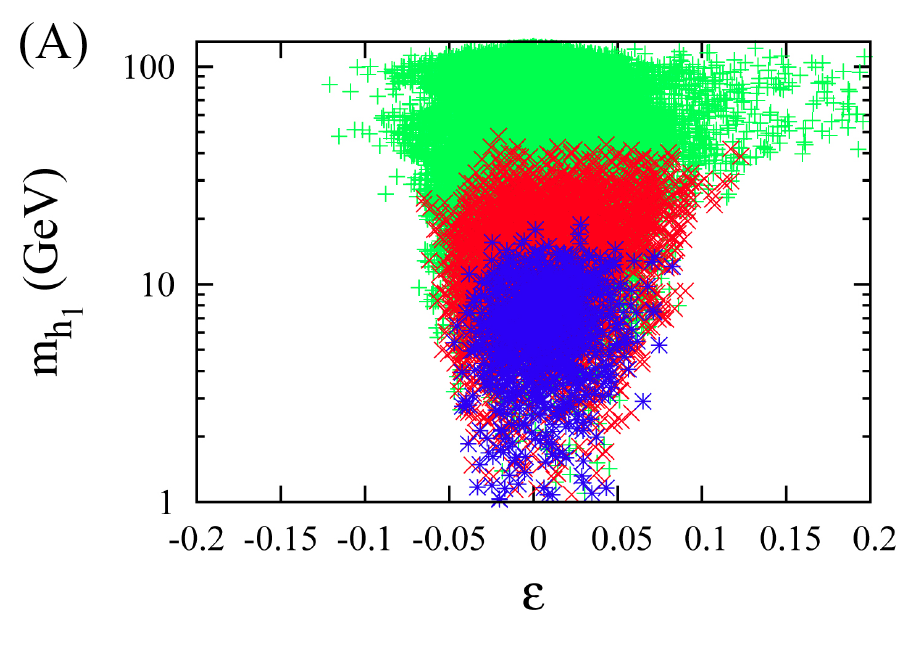}
\includegraphics[width=0.24\textwidth]{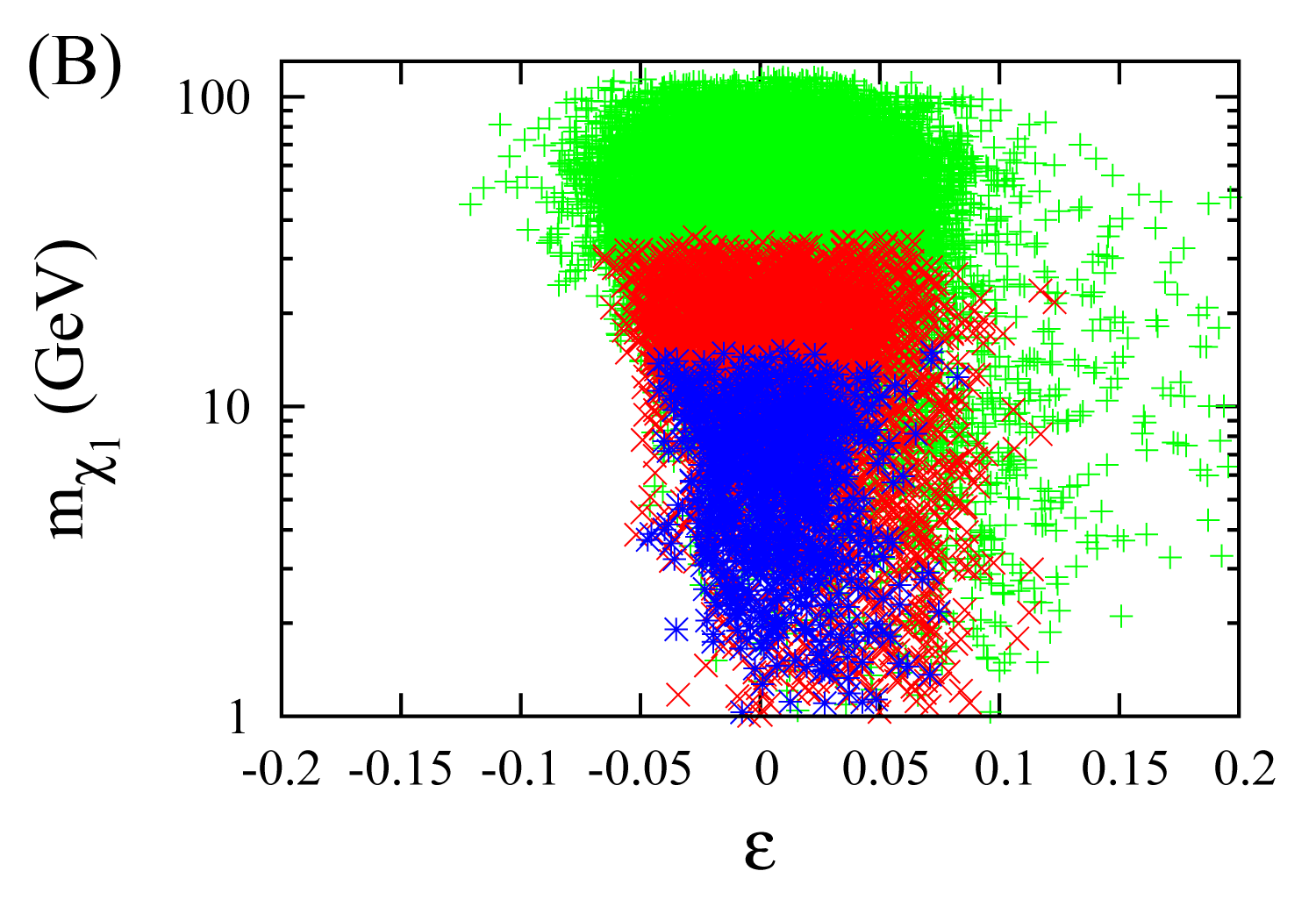}
\includegraphics[width=0.24\textwidth]{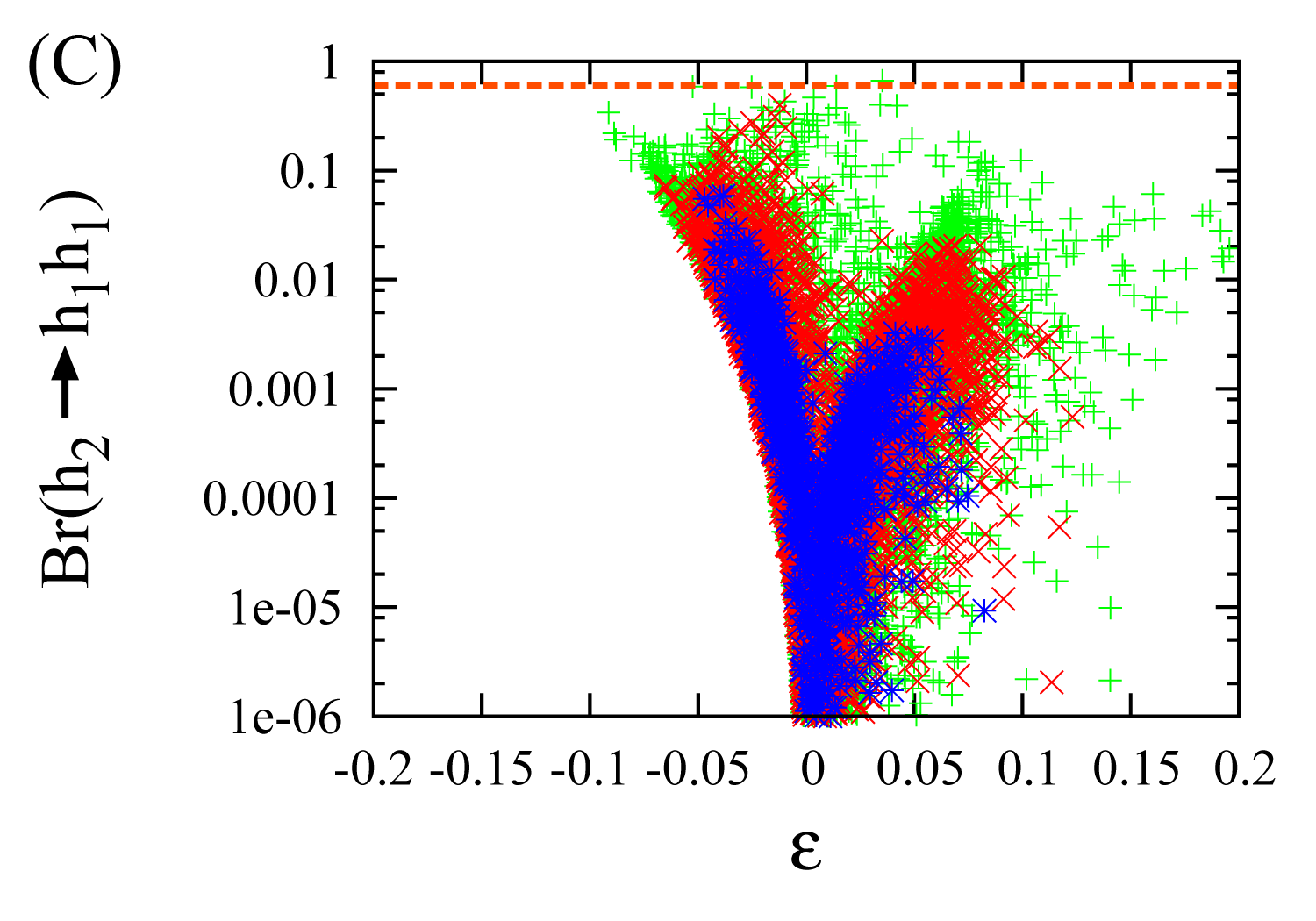}
\includegraphics[width=0.24\textwidth]{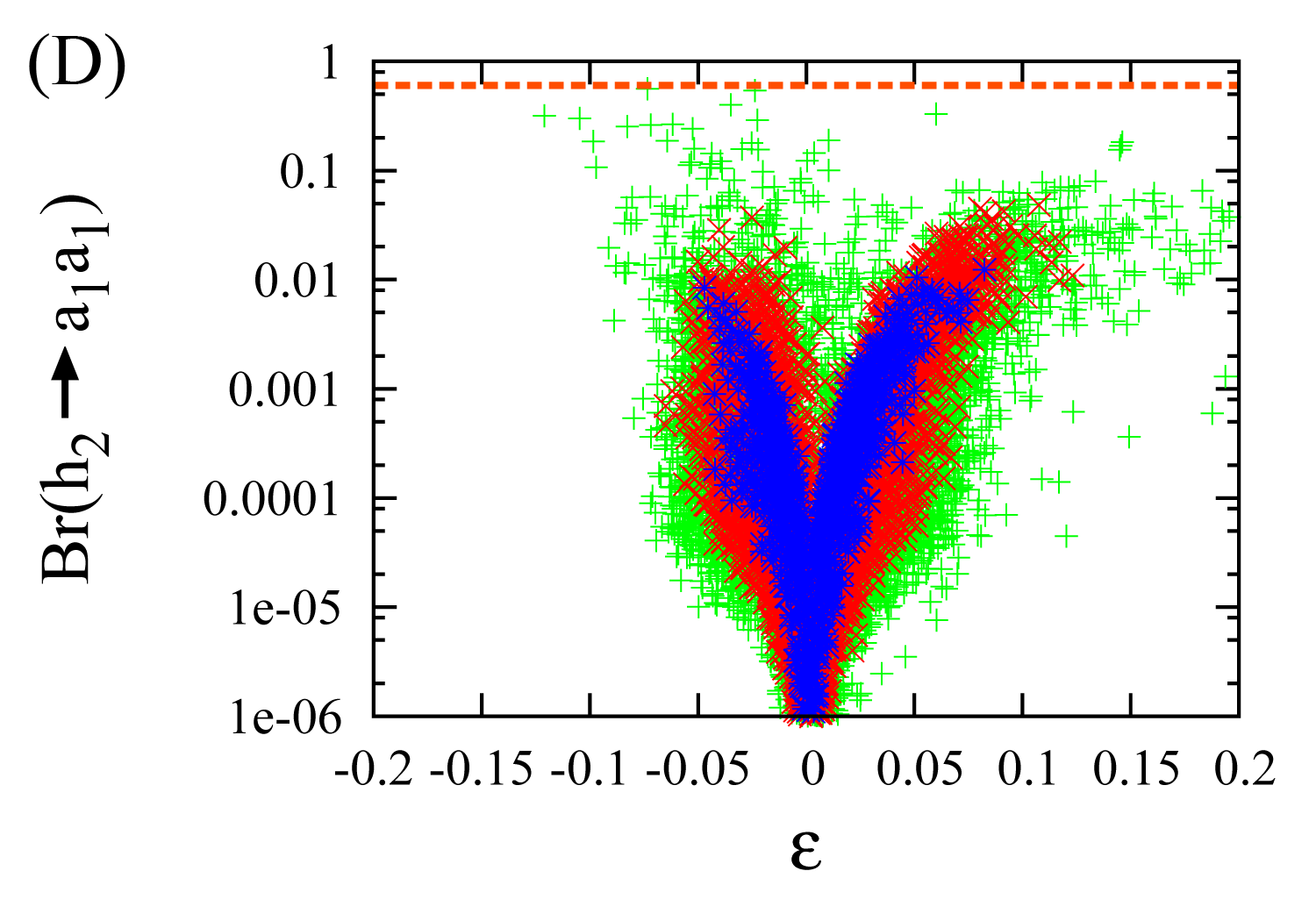}
\includegraphics[width=0.24\textwidth]{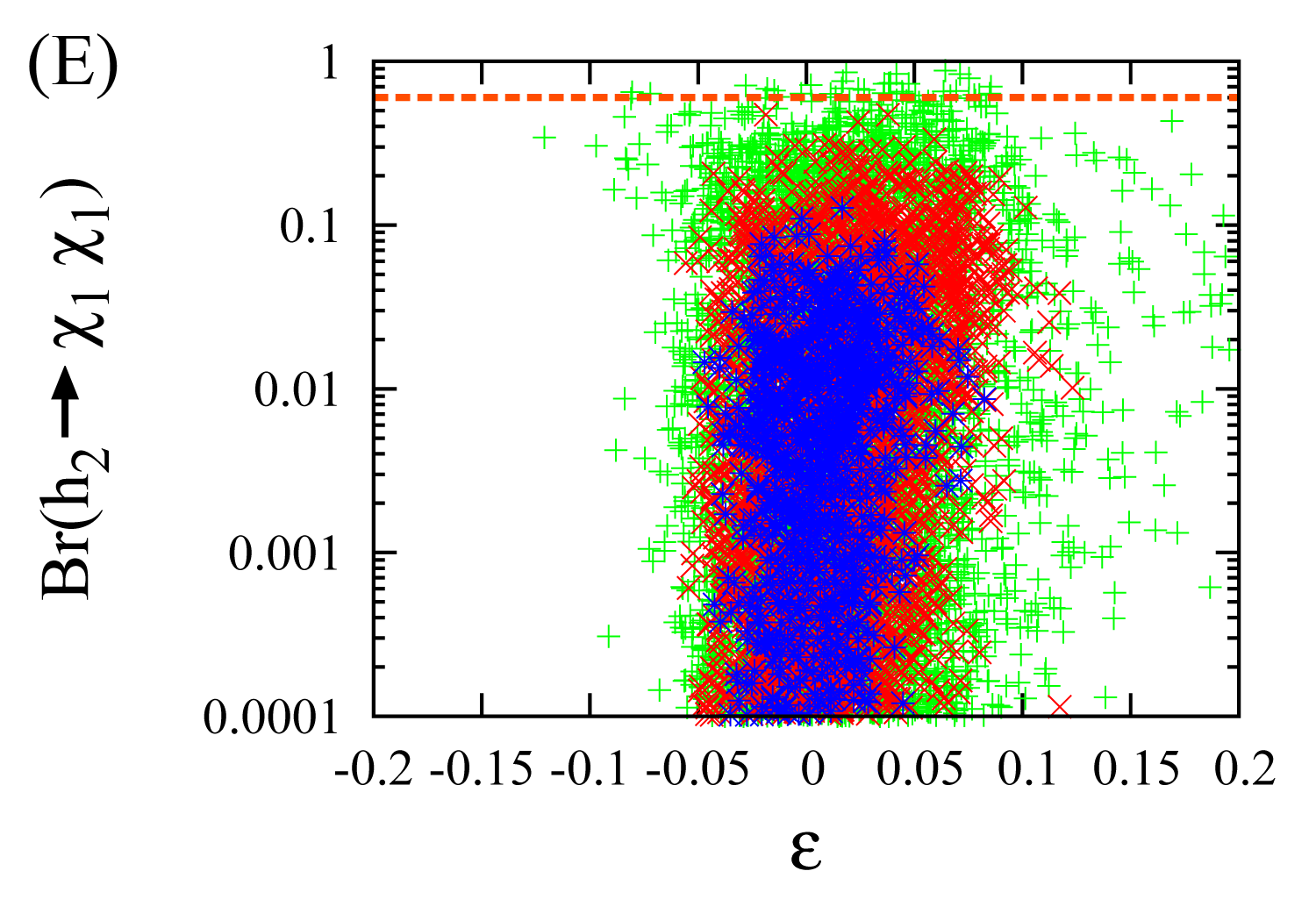}
\includegraphics[width=0.24\textwidth]{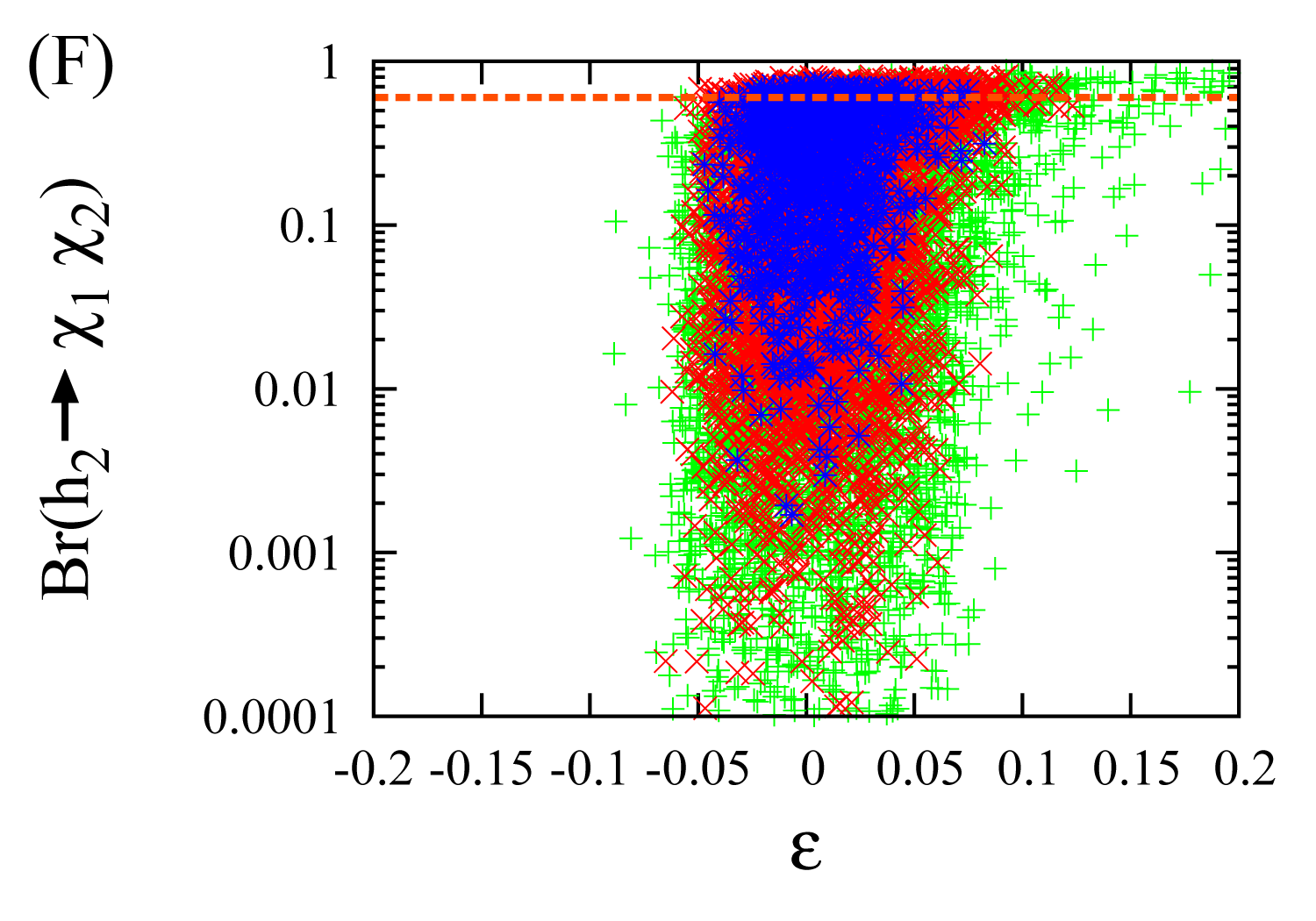}
\includegraphics[width=0.24\textwidth]{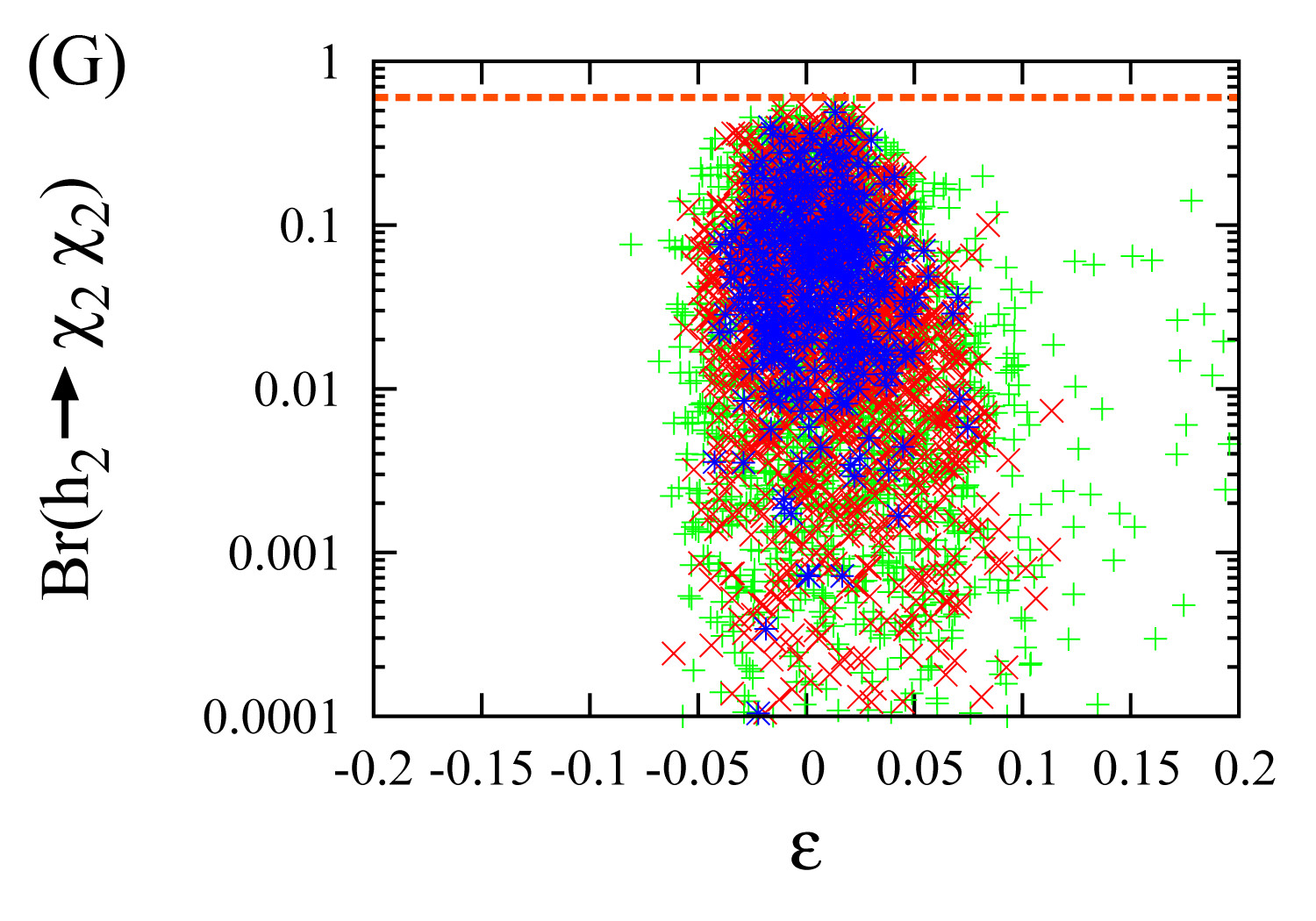}
\includegraphics[width=0.24\textwidth]{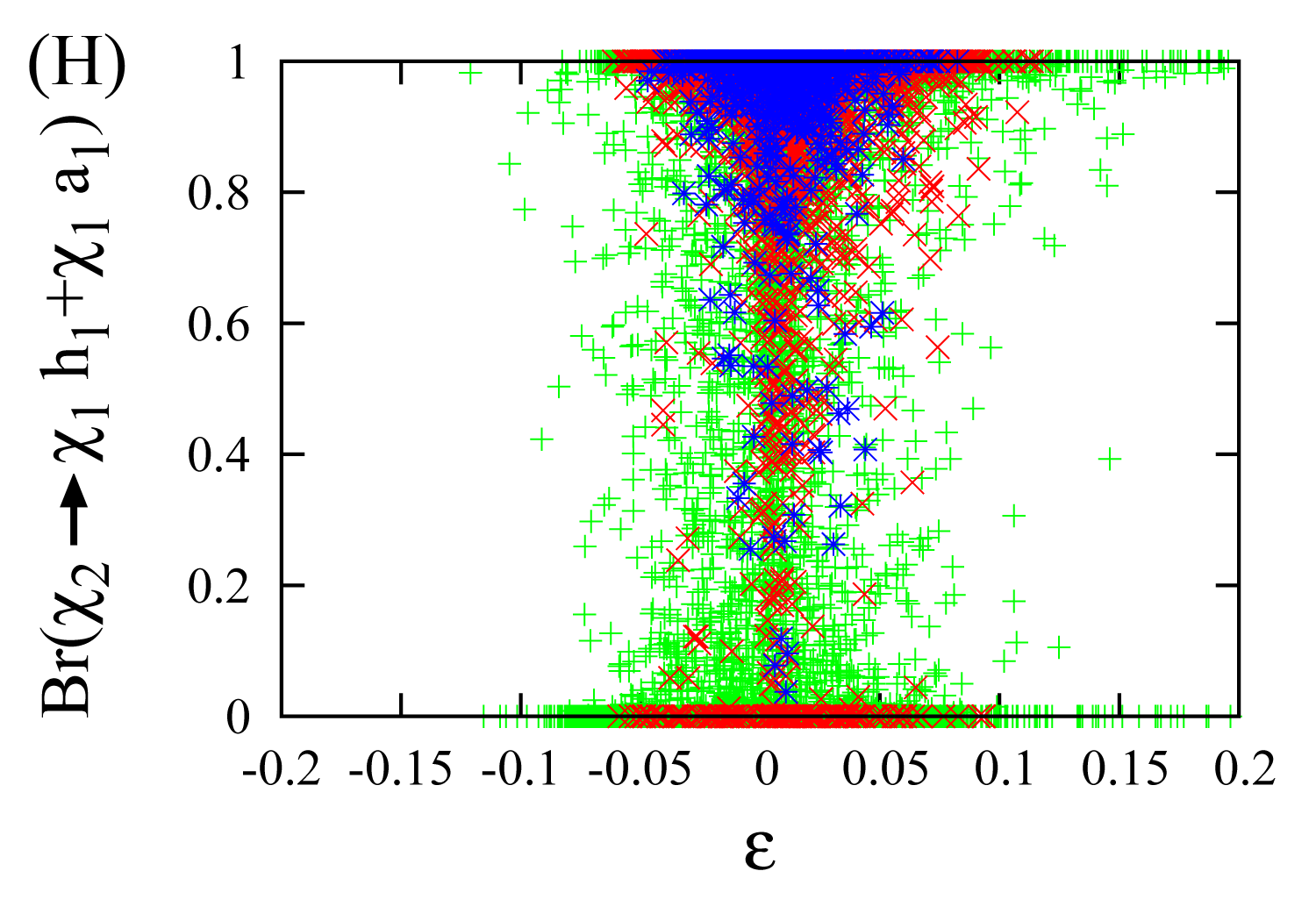}
\caption{Masses of (A) $h_1$ and (B) $\chi_1$; branching ratios of
  $h_2$ into (C) $h_1h_1$, (D) $a_1 a_1$, (E) $\chi_1\chi_1$, (F)
  $\chi_1 \chi_2$, (G) $\chi_2 \chi_2$ and (H) $\chi_2$ into $\chi_1
  h_1$ and $\chi_1 a_1$, in the PQ-symmetry limit of the NMSSM.
  Points are sampled in the ranges $3 \leq \tan \beta \leq 30$, $0.015
  \leq \lambda \leq 0.5$, $0.0005 \leq \kappa \leq 0.05$, $-0.8 \leq
  \varepsilon' \leq 0.8$, $-50 \text{ GeV} \leq A_\kappa \leq 0$, and
  $0.1 \text{ TeV} \leq \mu \leq 1 \text{ TeV}$.  We have assumed soft
  squark masses of 2 TeV, slepton masses of 200 GeV, $A_{u,d,e} =
  -3.5$ TeV, and bino, wino and gluino masses of 30-120, 150-500 and
  2000 GeV, respectively. Green (light gray) points cover the whole
  scan range, red (medium gray) points correspond to $\lambda < 0.30$,
  $\kappa / \lambda < 0.05$ and $\mu < 350$ GeV, and blue (dark gray)
  points correspond to $\lambda < 0.15$, $\kappa / \lambda < 0.03$ and
  $\mu < 250$ GeV.  In addition to the regular LEP, Tevatron and LHC
  bounds set in NMSSMTools 3.1.0~\cite{NMSSMTools}, we require
  $m_{h_2}$ to be within $124-126$ GeV. The dashed orange line in
  (C-G) depicts a $60\%$ exotic Higgs decay limit allowed at 2$\sigma$
  C.L.~\cite{Belanger:2013kya, Giardino:2013bma, Djouadi:2013qya,
    ATLAS:2013sla, CMS:yva} (see main text).}
\label{fig:pheno}
\end{center}
\vglue 0.5cm
\end{figure*}

Though kinematically allowed, the decays of the SM-like Higgs 
$h_2\to a_1a_1$, $h_1h_1$ are suppressed in the PQ-symmetry limit. 
The tree-level couplings of the SM-like Higgs boson $h_2$ with $h_1 h_1$ and $a_1
a_1$ are
\begin{eqnarray}
y_{h_2 a_1 a_1} &=& -\sqrt{2} \lambda\varepsilon \frac{m_Z v}{\mu}  + 
\sum\limits_{i=0}^{4} \mathcal{O} \left( \frac{\lambda^{4-i}}{\tan^i \beta}, \kappa \right) 
,  \\ 
y_{h_2 h_1 h_1} &=& -\sqrt{2} \lambda \varepsilon \frac{m_Z v}{\mu}  + 
2 \sqrt{2}   \varepsilon^2 v  + \sum\limits_{i=0}^{4} \mathcal{O} \left( 
\frac{\lambda^{4-i}}{\tan^i \beta}, \kappa \right) \nonumber ,
\label{eqn:yh2a1a1}
\end{eqnarray}
both of which are suppressed by $|\lambda \varepsilon| \ll 1$.  Unlike
the $R$-symmetry limit of the NMSSM (defined by $A_{\lambda,\kappa}
\to 0$), these decay channels are therefore rather inconsequential for
the SM-like Higgs in our scenario.

\begin{figure}[ht]
\begin{center}
\includegraphics[width=0.4\textwidth]{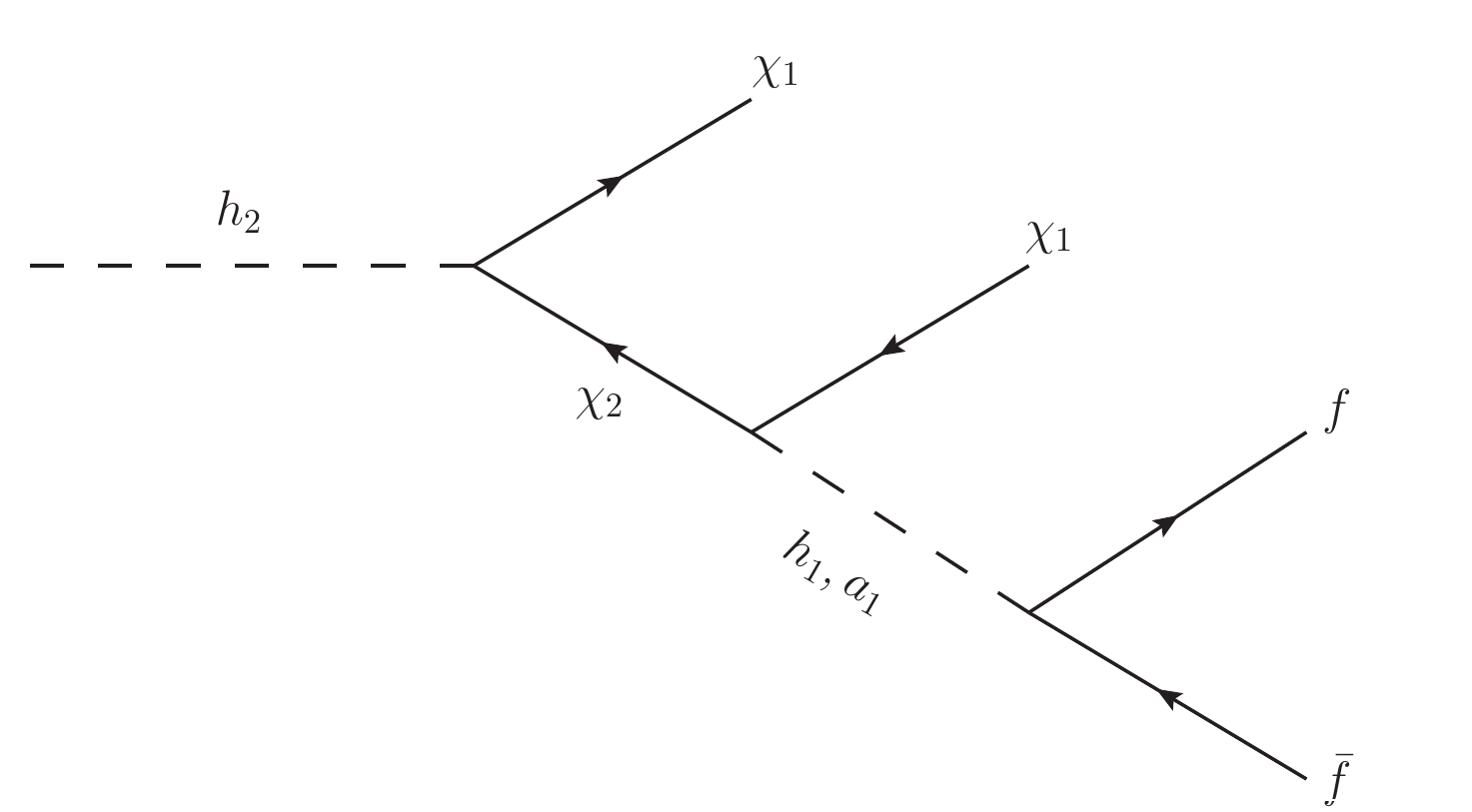}
\caption{A new decay topology of the SM-like Higgs boson.}
\label{fig:decay_topo}
\end{center}
\end{figure}

In the PQ-symmetry limit, however, the SM-like Higgs has a significant
decay width into a pair of neutralinos $h_2 \to \chi_1 \chi_2$,
$\chi_2 \chi_2$, if kinematically allowed.  Since $\chi_1$ is light,
$h_2 \to \chi_1 \chi_2$ can be significant, {\it e.g.}, if $\chi_2$ is
bino-like, with $m_{\chi_2} \lesssim 100$ GeV (note $h_2 \to \chi_2
\chi_2$ is also possible but tends to be phase-space suppressed),
while $h_2 \to \chi_1\chi_1$ is suppressed by the small mixing angle
of the singlino-like $\chi_1$.  Their relative strength can be
understood via the couplings
\begin{eqnarray}
y_{h_2 \chi_1 \chi_2} \sim \mathcal{O} \left(\frac{\lambda g_1 v}{ \mu}
\right) \ , \quad
y_{h_2 \chi_1 \chi_1} \sim \mathcal{O} \left(
\frac{\lambda^2 v}{\mu \tan \beta} \right) \ .
\end{eqnarray}
The decay width $\Gamma_{h_2 \to \chi_1 \chi_2}$ therefore is
typically larger than $\Gamma_{h_2 \to \chi_1 \chi_1}$.  Given that
$\Gamma_{h \to b \bar{b}}$ is dictated by the coupling $\frac{m_b}
{\sqrt{2} v}$, Br$(h_2 \to \chi_1 \chi_2)$ can be sizable, as shown in
Fig.~\ref{fig:pheno}, and even larger than the partial widths to SM
final states.  The $\chi_2$ dominantly decays into $\chi_1 a_1$ or
$\chi_1 h_1$, which are usually the only kinematically accessible
channels.  Therefore, the PQ-symmetry limit of the MSSM
singlet-extensions provides new supersymmetric decay topologies for
the SM-like Higgs boson, including the one shown in
Fig.~\ref{fig:decay_topo}, which we now discuss in detail.

{\bf [LHC Studies]} The branching fractions of $h_1$ and $a_1$ (we do
not distinguish $h_1$ and $a_1$ below) into the SM fermions $f \bar f$
are highly sensitive to their masses.  The decay products $f \bar f$
tend to be soft, because of the restricted phase space, and to be 
collimated: their characteristic angular separation $\Delta R \equiv
\sqrt{ (\Delta \eta)^2 + (\Delta\phi)^2 } \sim 2 m_{h_1} / p_{T, h_1}$
is small since typically $m_{h_1} \ll p_{T, h_1}$, where $\Delta \eta$
and $\Delta \phi$ are the differences in pseudorapidity and azimuthal
angle of $f \bar f$, respectively.  Our new exotic Higgs decay
collider signature is therefore $f \bar{f} + \met + X$, where $f\bar
f$ behaves as lepton-jet(s) or jet(s), and $X$ denotes the particles
associated with the $h_2$ production.  Two benchmarks, with
$f\bar f = \mu^+ \mu^-$ and $b\bar b$, respectively, are listed in
Table~\ref{table:benchmark}. Another interesting possibility is $f\bar
f = \tau^+\tau^-$ which is challenging, however,
because of the failure of the standard tau identification method due
to the softness and collimation of the $\tau^+\tau^-$ signal.  Thus,
extracting the ditau signal requires a sophisticated treatment of
backgrounds, exemplified by QCD and soft tracks.  We will present the
relevant sensitivity analysis in~\cite{HLWY}.

\begin{table} [ht]
\begin{tabular}{c|c|c|c|c}\hline\hline
&$m_{h_1}$&$m_{h_2}$& $m_{\chi_1}$ &  $m_{\chi_2}$  \\ \hline
$h_1 \to \mu^+\mu^-$ & 1 GeV& 125 GeV & 10 GeV & 80 GeV   \\ \cline{1-2}
$h_1 \to b \overline{b}$  & 45 GeV& & & \\ \hline
\end{tabular}
\caption{Two benchmarks used for the collider analyses.}
\label{table:benchmark}
\end{table} 

To design the collider strategies which can optimally cover
the full space of models, we perform the analysis model-independently 
and introduce a rescaling factor
\begin{eqnarray}
c_{\text{eff}} & = & \frac{\sigma(pp \to h_2)}{\sigma(pp \to h_{\rm SM})} 
\times {\rm Br}(h_2 \to \chi_1 \chi_2) \\ \nonumber
& & \times {\rm Br}(\chi_2 \to h_1 \chi_1) 
\times {\rm Br}(h_1 \to f \bar{f}) \ ,
\end{eqnarray}
where $\sigma(pp \to h_2, h_{\text{SM}})$ are the production cross
sections for the SM-like and SM Higgs in the relevant production mode,
and we assume the narrow width approximation for each intermediate
decaying particle.  Given the absence of dedicated collider searches
so far, the current upper bound for $c_{\rm eff}$ is from fitting the
results of standard Higgs searches, and hence is not sensitive to
$m_{h_1}$.

The samples for both analyses are simulated using MadGraph
5~\cite{Alwall:2011uj} with CTEQ6L1 parton distribution
functions~\cite{Pumplin:2002vw} and, for the $h_1 \to b\bar b$ benchmark,
MLM matching~\cite{Mangano:2001xp, Mangano:2002ea}.  They are showered
and hadronized using \textsc{Pythia} v6.4.20~\cite{Sjostrand:2006za}.
For the $\mu^+ \mu^-$ case, we use PGS v4~\cite{PGS4} for basic
detector simulation.  For the $b \bar{b}$ case, since we will perform
a jet substructure analysis, we use a more sophisticated mock detector
simulation based on physics object studies performed by ATLAS and CMS
for jets~\cite{Chatrchyan:2011ds}, electrons~\cite{Aad:2011mk},
muons~\cite{ATLAS:2011hga}, and $\met$~\cite{Chatrchyan:2011tn}.

{\bf [Case I: $h_2\to \mu^+\mu^- +\met$]} For $m_{h_1} \lesssim 1$
GeV, the dominant decay channel is $h_1\to \mu^+\mu^-$, resulting in
$\Delta R_{\mu^+\mu^-} \sim 0.1$ for the benchmark point in
Table~\ref{table:benchmark}.  These muons generally fail the usual
isolation requirements in multilepton SUSY searches (where summing
over all particles' $p_T$ in the cone around each muon is typically
assumed), rendering such searches insensitive to this channel.  In
addition, though searches for lepton-jets at the LHC~\cite{Aad:2012kw,
  Aad:2012qua, Aad:2013yqp} do not impose isolation requirements on
the collimated leptons, they make additional requirements which render
them insensitive to our model.  Namely, these searches require a
displaced vertex for the lepton-jet~\cite{Aad:2012kw}, at least four
muons within a single lepton-jet~\cite{Aad:2012qua}, or at least two
lepton-jets~\cite{Aad:2012qua, Aad:2013yqp}: all of these features are
absent in this scenario.  The most sensitive search comes from
Ref.~\cite{Chatrchyan:2011hr}, which used 35 pb$^{-1}$ of 7 TeV data
to search for resonances decaying to muon pairs.  After applying their
analysis cuts, we obtain a signal cross section of $\sigma (gg \to
h)_{SM} \times c_{\text{eff}} \times A \sim 0.1 \text{ pb} \times
c_{\text{eff}}$, which well satisfies their $0.15-0.7$ pb limit for
masses below 1 GeV~\cite{Chatrchyan:2011hr}.

We now proceed with developing a collider analysis for identifying the
dimuon signal in our model.  As an illustration, we focus on the $W
h_2$ mode with the $W$ decaying leptonically ($\ell = e$, $\mu$).  The
muons from the $h_1$ decay are collimated, and so we define a new muon
isolation cut, which requires the particle $p_T$ sum (excluding the
nearby-muon contribution) in the $\Delta R < 0.4$ cone around each
muon candidate to be $p_{T, iso} (\mu^{\pm}) < 5$ GeV.  This can
efficiently discriminate the $h_1$ muons and the ones from
semi-leptonic meson decays.  The main background then arises from $W
\gamma^*/Z^* \to \ell \nu \mu^+ \mu^-$.  We assume a tri-lepton
trigger in the analysis.  Alternatively, we could have triggered on
the single lepton from the $W$ decay, which would not significantly
alter our conclusions.
\begin{figure}[ht]
\begin{center}
\includegraphics[width=0.23\textwidth]{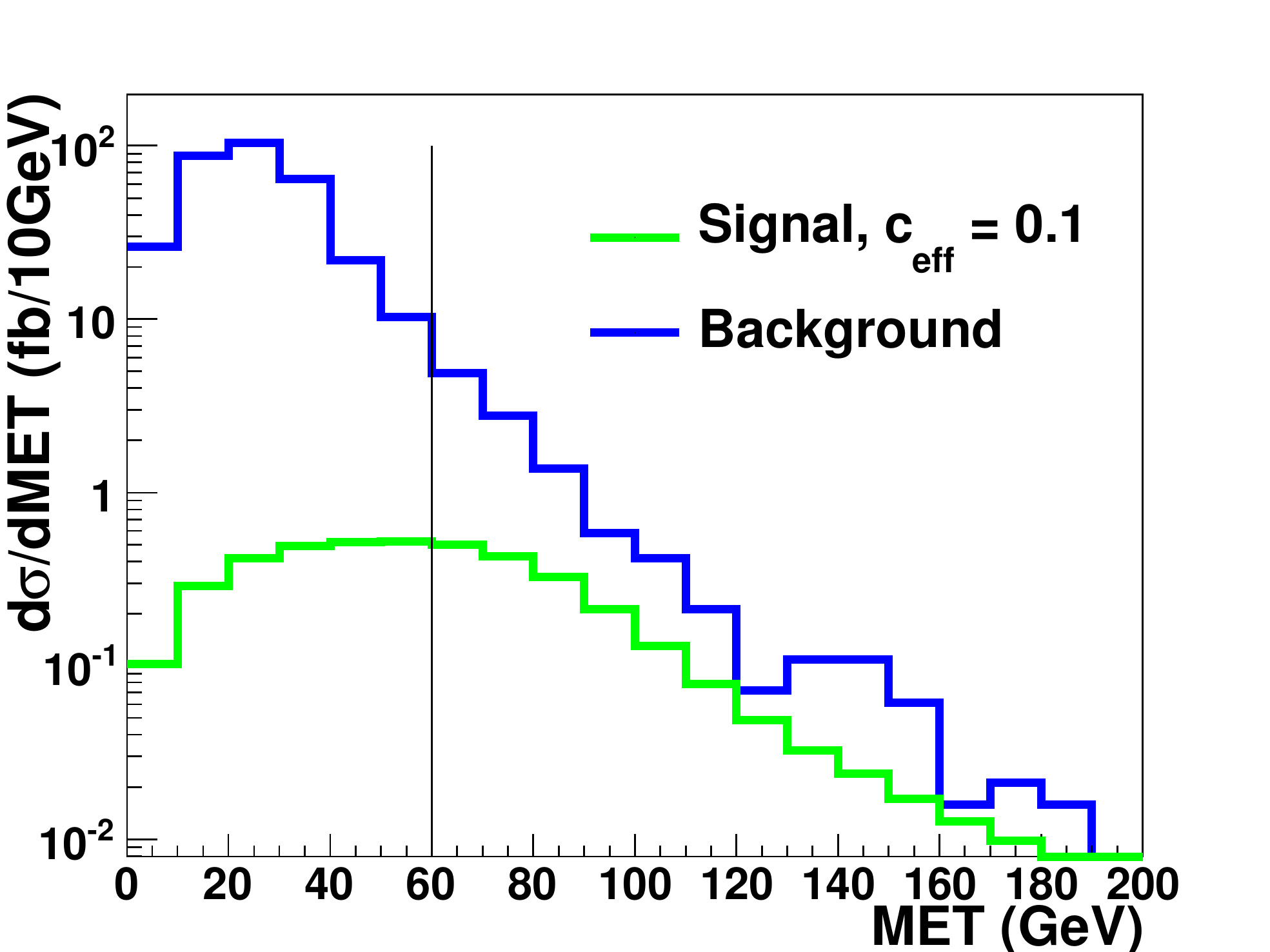}
\includegraphics[width=0.23\textwidth]{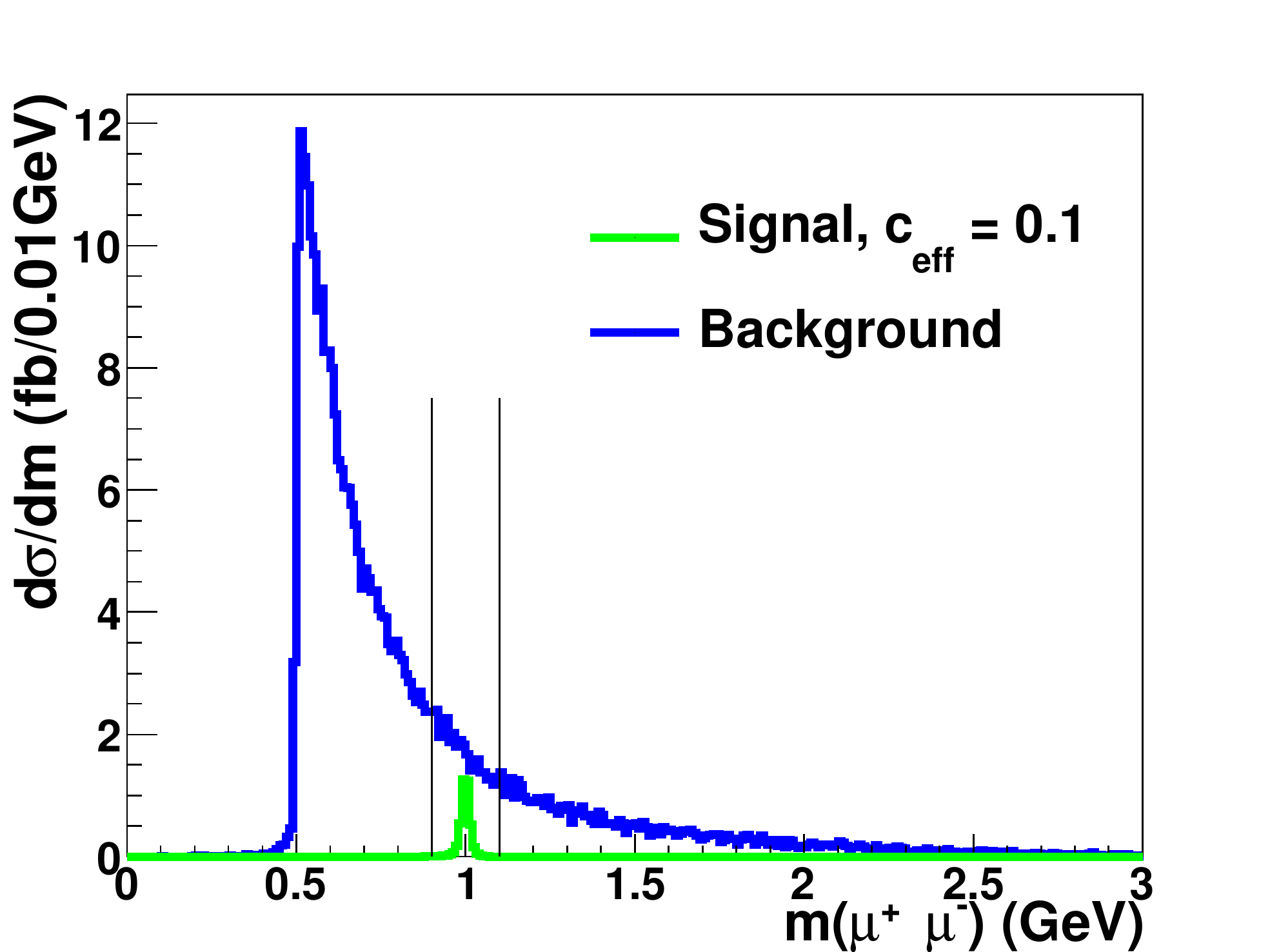}
\caption{Differential cross section vs. $\met$ (left) and
  $m_{\mu^+\mu^-}$ (right) at the 8 TeV LHC.  We have applied basic
  acceptance cuts to both signal (choosing $c_{\text{eff}} = 0.1$) and
  background and additional cut of $m_{\mu^+\mu^-} \geq 0.5$ GeV to
  the background to ensure efficient Monte Carlo (MC) generation.  The
  black line indicates the cuts $\met > 60$ GeV and $0.9 < m_{\mu^+
    \mu^-} < 1.1$ GeV, respectively. }
\label{fig:dis_dimu1}
\end{center}
\end{figure}

\begin{table}[ht]
\begin{tabular}{p{4cm}|c|c}\hline\hline
Cuts and Efficiencies & $W h_2$
& $W (\gamma^*/Z \to \mu^+\mu^-)$ \\ \hline
Cross section (pb) & $0.149 \times c_{\rm eff}$  & 26.6  \\ \hline
Lepton geometric, $p_T$, and isolation requirements
& 28.2\% & 1.22\% \\ \hline
$\met \ge$ 60 GeV                      & 12.5\% & 0.0403\% \\ \hline
0.9 $\le m_{\mu^+ \mu^-}\le$ 1.1 GeV       & 12.3\%  & 0.0047\% \\ \hline
$S$, $B$, $\frac{S}{\sqrt{B}}$ (20 fb$^{-1}$, $c_{\text{eff}} = 0.1$) 
& \multicolumn{2}{c}{37, 32, 6.5 $\sigma$} \\ \hline 
\end{tabular}
\caption{Analysis cuts (see main text) and efficiencies at the 8 TeV
  LHC for the $h_1 \rightarrow \mu^+\mu^-$ channel.  The decay $W \to
  \ell \nu$, $\ell = e$, $\mu$ is assumed in the quoted cross section,
  and a $K$-factor of 1.3 is included for the background.}
\label{table:WmMvH2DiMu115GeV}
\end{table} 

In Fig.~\ref{fig:dis_dimu1}, we show the $\met$ and $m_{\mu^+\mu^-}$
distributions after imposing basic acceptance cuts at the 8 TeV LHC.
Namely, we require $|\eta_{\ell}| \leq 2.4$ for all charged leptons,
$\Delta R_{\mu^+ \mu^-} < 0.2$ and $p_{T, \mu} \geq 10$ GeV for
candidate $h_1$ muon pairs, and $p_{T, \ell} \geq 20$ GeV for a third
lepton.  We also impose the new muon isolation requirement detailed
above.  The complete cut efficiency is presented in
Table~\ref{table:WmMvH2DiMu115GeV}.  With 20 fb$^{-1}$ data, we can
obtain a local statistical sensitivity $S/\sqrt{B} = 6.5 \sigma$, with
a typical value of $c_{\rm eff} = 0.1$ (see discussions, e.g.,
in~\cite{HHG,Curtin:2013fra}) assumed.  A larger significance is
certainly possible, though, if the $m_{\mu^+\mu^-}$ cut was tightened,
or additional triggers were added.  This dedicated analysis can be
easily extended to the other possibilities, {\it e.g.} $Z h_2$ events.
On the other hand, we have neglected the background resulting from
jets and photons faking electrons and lost jets faking $\met$, as well
as subleading sources of fake muons.  We expect all of these
contributions arising from fakes to be subdominant compared to the
irreducible $W \gamma^* / Z^*$ background.


{\bf [Case II: $h_2\to b \bar b +\met$]} For $m_{h_1} > 10$ GeV, $h_1$
dominantly decays into a relatively soft $b \bar{b}$ pair.  Unlike
previous Higgs boson studies in the $b \bar{b}$
channel~\cite{Butterworth:2008iy, Kribs:2009yh, Kribs:2010hp}, the
$h_1$ boson in our scenario results from the cascade decay of a
$\mathcal{O} (100)$ GeV parent particle, imparting only a mild boost
to the $b \bar{b}$ system.  We will reconstruct the $h_1 \rightarrow b
\bar{b}$ signal using jet substructure tools appropriately modified
for non-boosted resonances.  To avoid washing out our hadronic signal
and because our $\met$ significance is not strong enough for a
standalone trigger, we will focus on the $Z h_2$ mode and trigger on
the leptonic $Z$ decay.  Then the dominant background is $Z + $ heavy
flavor jets.

We first aim to identify the $Z$ candidate from its same flavor
opposite sign (SFOS) $e^+ e^-$ or $\mu^+ \mu^-$ decays, which can
efficiently remove the $t \bar t$ background, then we find the
$b$-tagged $h_1$ candidate jet and probe its substructure.  We cluster
jets using the angular-ordered Cambridge-Aachen
algorithm~\cite{Dokshitzer:1997in, Wobisch:1998wt} from
\textsc{FastJet} v3.0.2~\cite{Cacciari:2011ma} with distance parameter
$R = 1.2$. After applying the $\met > 80$ GeV cut, we count the number
of $b$-tagged jets with $p_T > 20$ GeV.  Our $b$-tagging efficiency is
60\% with a mistag rate 10\% for $c$-jets and $1\%$ for the other
light jets. We choose to retain the 1 $b$-tag bin for the jet
substructure analysis.

\begin{figure}[ht]
\begin{center}
\includegraphics[width=0.23\textwidth]{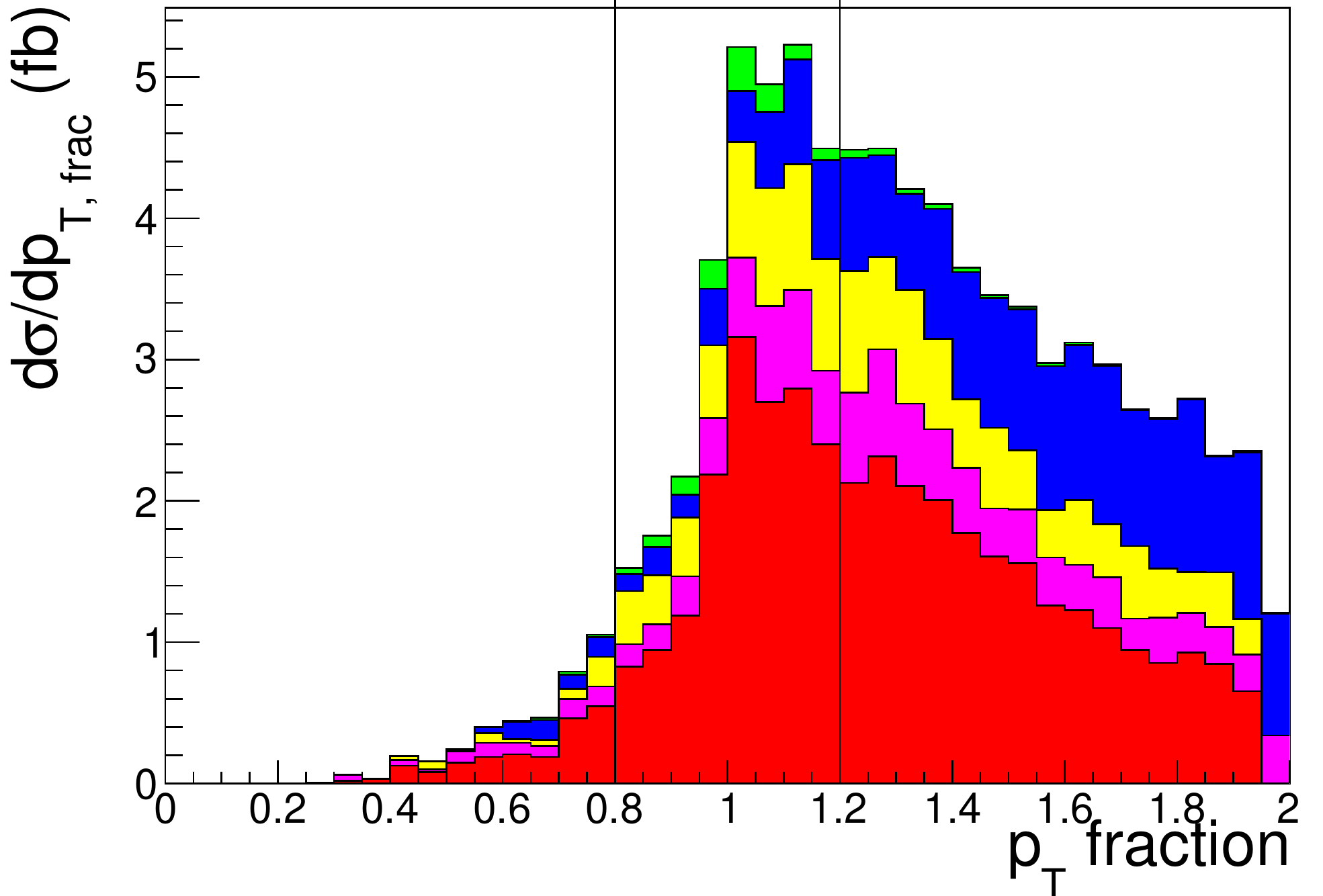}
\includegraphics[width=0.23\textwidth]{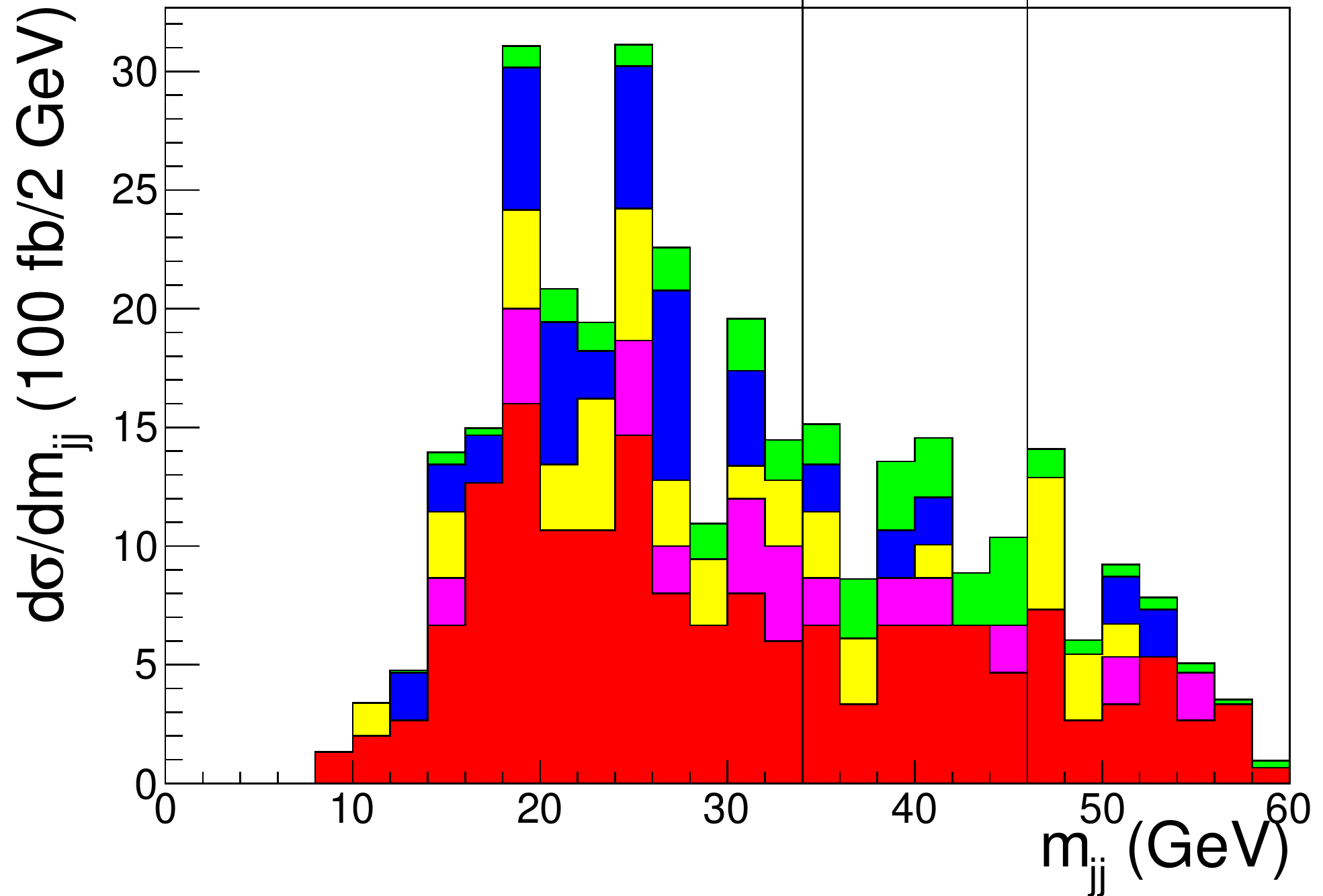}
\caption{Differential cross section vs. $p_{T, {\rm frac}}$ (left) and
  $m_{jj}$ (right), as defined in the text, after applying the $Z$
  mass window, $\met$, and $N_{b-\text{tag}} = 1$ cuts (and $p_{T,
    {\rm frac}}$ cut) at the 14 TeV LHC.  Red is $Zb \bar b$
  background, blue is $t\bar t$ background, magenta is the $Zc \bar c$
  background, yellow is the $Zc + Z \bar c$ background and green is $Z
  h_2$ signal.  The black vertical lines indicate the $p_{T, {\rm
      frac}}$ window, $0.8 < p_{T, \text{frac}} < 1.2$, and the
  $m_{jj}$ window, requirement in our analysis.  We have set
  $c_{\text{eff}} = 0.5$.}
\label{fig:bb_pTmatch}
\end{center}
\end{figure}

Having isolated the dilepton system as well as the cascade decay of
the $h_2$ boson, we can apply an additional cut with the expectation
that the dilepton system recoils against the collimated $h_2$ cascade
decay.  We construct the scalar sum $p_T$ of the $h_2$ candidate, $p_T
(h_2) = p_T (b-\text{jet}) + |\met|$, and then divide it by
the $p_T$ of the dilepton system: $p_{T, \text{ frac}} \equiv p_T
(h_2) / p_T (\ell \ell_{\text{sys}})$.  The $p_{T, \text{ frac}}$
distribution is shown in Fig.~\ref{fig:bb_pTmatch}.  We observe that
the cutting on $p_{T, \text{ frac}}$ works well at reducing the $t
\bar{t}$ background, where the $\met$ signal tends to arise from
neutrinos of separate decay chains instead of a single cascade decay.

\begin{table}[ht]
\scalebox{0.9}{
\begin{tabular}{p{3cm}|c|c|c|c|c}\hline \hline
Cut, efficiencies  & $Z h_2$ & $Z b \bar b$ & $Z c \bar c$ & $Z c + Z \bar c$ & $t \bar{t}$ \\ \hline
Cross section (pb) & 0.09$\times c_{\text{eff}}$ 
                   & 48.4    & 32.8 & 139 & 41.8          \\
\hline 
Lepton cut     &  
0.191 &  0.177 & 0.170 & 0.163 & 0.012 \\ \hline
$\met > 80$ GeV                         & 8.31E-2
                                        & 3.15E-3
                                        & 4.41E-3
                                        & 1.80E-3
                                        & 4.57E-3 \\ \hline
$N_{b-\text{tags}} = 1$, jet $p_T > 20$ GeV & 3.78E-2
                                        & 1.09E-3
                                        & 4.88E-4
                                        & 1.31E-4
                                        & 1.86E-3 \\ \hline
$0.8 < p_{T, \text{ frac}} < 1.2$           & 2.59E-2
                                        & 3.35E-4
                                        & 1.06E-4
                                        & 3.59E-5
                                        & 7.70E-5 \\ \hline
Two hardest subjets: $34 < m_{jj} < 46 {\rm \ GeV}$  
                                        & 3.48E-3
                                        & 7.17E-6
                                        & 2.44E-6
                                        & 5.00E-7
                                        & 1.44E-6 \\ \hline
\hline
$S, B, \frac{S}{\sqrt{B}}$ (600 fb$^{-1}$, $c_{\text{eff}} = 0.5$)
                                        & 
\multicolumn{3}{c}{93, 208+48+42+36,  5.1 $\sigma$} \\ 
\hline 
\end{tabular}}
\caption{Analysis cuts and efficiencies at 14 TeV LHC for the $h_1
  \rightarrow b \bar{b}$ channel. The standard cross section
  normalizations have been adopted: $833$ pb for $t\bar t$
  events~\cite{Bonciani:1998vc}, and $0.883$ pb for $Zh_2$
  events~\cite{Dittmaier:2011ti}. A $K$-factor of 1.3 is included for
  the other backgrounds. The decays $Z \to \ell^+ \ell^-$ and, for top
  decays, $W \to \ell \nu$, $\ell = e$, $\mu$, $\tau$ are included in
  the quoted cross sections, where preselection cuts have been
  applied. The lepton cut requires the two hardest leptons ($\ell = e,
  \mu$) satisfy: SFOS, $p_T > 40$ GeV, $|m_{\ell \ell} - m_Z| < 10$
  GeV.  }
\label{table:bb}
\end{table}

Finally, we apply jet substructure techniques to investigate the
kinematics of the $b$-tagged, $R = 1.2$ jet.  Since we expect the $Z +
$ heavy flavor jets background to include more final state radiation
than our $h_1 \rightarrow b \bar{b}$ signal, we recluster the $R =
1.2$ jet using a smaller cone size of $R = 0.3$. We count the number
of subjets with $p_T > 10$ GeV, and for events with two such subjets,
we plot the invariant mass of the subjets in
Fig.~\ref{fig:bb_pTmatch}.  We can readily observe a feature close to
the $m_{h_1} = 45$ GeV mass in the data.  Complete cut flow
information and sensitivity calculation are presented in
Table~\ref{table:bb}.  With 600 fb$^{-1}$ of 14 TeV LHC data, in the
mass window of $34 < m_{jj} < 46$ GeV, the excess has a local
significance of $S / \sqrt{B} = 5.1\sigma$, with $c_{\text{eff}} =
0.5$ assumed~\footnote{A $c_{\text{eff}}$ value as large as 0.5
  usually requires new physics enter the Higgs-gluon-gluon coupling,
  otherwise the upper bound is about 0.25 at 95\%
  C.L.~\cite{Belanger:2013kya, Giardino:2013bma, Cao:2013gba}.  While
  our model may potentially accommodate this requirement, studying the
  relevant physics is beyond the scope of this letter.  Instead, we
  remark that the sensitivity for different $c_{\text{eff}}$ values
  can be easily obtained via rescaling.  Moreover, additional
  production modes for the Higgs, such as supersymmetric cascade
  decays, can have kinematics that readily overlap with our benchmark
  modes and effectively enhance $c_{\text{eff}}$.}.  The sensitivity
could potentially be improved by further refining jet substructure
techniques to tease out the soft $h_1$ signal subjets from the
difficult hadronic collider environment, or choosing a new trigger
such as vector boson fusion.


{\bf [Summary]} In summary, the 125 GeV Higgs may be the leading
window into NP, while its exotic decays provide a very natural and
efficient way to explore such NP.  The PQ-symmetry limit of the MSSM
singlet-extensions provides a supersymmetric benchmark not only for a
DM candidate at the sub-EW scale, but also for a third category of
exotic decays of the 125 GeV Higgs (including both $h_2 \to
\chi_1\chi_2$ and $h_2 \to \chi_2\chi_2$). This category of exotic
Higgs decays are characterized by novel collider signatures - visible
objects ($b\bar b, \tau^+\tau^-, l^+l^-$, lepton jets, $\gamma$, etc.)
+ $\met$, and have rarely been considered before (though similar
topologies may also arise in some neutrino models where the Higgs
decays into two different neutrinos with the heavier one further
decaying via an off-shell gauge boson, e.g.,
see~\cite{Graesser:2007yj, Kersten:2007vk, de Gouvea:2007uz}; for
different topologies with a fermions (jets)+$\met$ signature, which may 
occur, e.g., in the multi-Higgs models with extra singlet,
see~\cite{Curtin:2013fra} (\cite{Englert:2012wf})).  This motivates
new directions in exploring exotic Higgs decays at colliders, and also
opens a new avenue to probe for new physics couplings with the 125 GeV
Higgs boson.

Indeed, given their significant role as new physics probes, a
systematic survey exploring exotic Higgs decays (including the other
categories of collider signatures: (1) $h\to$ purely $\met$ and (2)
$h\to$ visible objects) was pursued by ref~\cite{Curtin:2013fra}. The
exotic Higgs decays with various topologies and collider signatures
are prioritized according to their theoretical motivation and their
experimental feasibility.  Some highly motivated searches for the LHC
are the channels proposed in this letter, $h_2\to \chi_1 \chi_2,
\chi_2\chi_2$ with $\chi_2$ further decaying leptonically.  For more
challenging cases, suggestions for improving their search
sensitivities, i.e. via new triggers, are made.  For more details in
this regard, please see ref~\cite{Curtin:2013fra} or visit the website
\url{http://exotichiggs.physics.sunysb.edu/web/wopr/}.

\begin{center}
{\bf Acknowledgments}
\end{center}  

We would like to thank Brock Tweedie, Patrick Draper, Michael
Graesser, Joe Lykken, Adam Martin, Jessie Shelton, Matt Strassler and
Carlos Wagner for useful discussions.  TL is supported by his start-up
fund at the Hong Kong University of Science and Technology.  JH is
supported by the DOE Office of Science and the LANL LDRD program.  JH
would also like to thank the hospitality of University of Washington,
where part of the work was finished.  L-TW is supported by the DOE
Early Career Award under Grant DE-SC0003930.  L-TW is also supported
in part by the Kavli Institute for Cosmological Physics at the
University of Chicago through NSF Grant PHY-1125897 and an endowment
from the Kavli Foundation and its founder Fred Kavli.  Fermilab is
operated by the Fermi Research Alliance, LLC under Contract
No. DE-AC02-07CH11359 with the US Department of Energy.

We also would like to acknowledge the hospitality of the Kavli
Institute for Theoretical Physics and the Aspen Center for Physics, 
where part of this work was completed, and this research is supported 
in part by the National Science Foundation under Grant No. NSF PHY11-25915.


\end{document}